\documentclass[a4paper,12pt]{article}
\pdfoutput=1

\usepackage{graphicx}
\usepackage{amssymb}
\usepackage{cite}
\usepackage{bm}
\usepackage{indentfirst}
\usepackage{amsmath}
\usepackage{hhline}
\usepackage{multirow}
\usepackage{slashed}
\usepackage{hyperref}
\hypersetup{colorlinks,bookmarksopen,bookmarksnumbered,citecolor=blue,
linkcolor=black,pdfstartview=FitH,urlcolor=blue}
\usepackage{ulem} 
\usepackage{verbatim}
\newcommand{\MeV}[0]{\,\mathrm{MeV}}
\newcommand{\GeV}[0]{\,\mathrm{GeV}}

\begin{document}

\begin{titlepage}

\begin{flushright}
TUM-HEP-1129-18 \\ KIAS-P18007\\
\end{flushright}

\begin{center}

\vspace{1cm}
{\large\bf 
A new mechanism of sterile neutrino dark matter production}
\vspace{1cm}

\renewcommand{\thefootnote}{\fnsymbol{footnote}}
Johannes Herms$^1$\footnote[1]{johannes.herms@tum.de}
,
Alejandro Ibarra$^{1,2}$\footnote[2]{ibarra@tum.de}
,
Takashi Toma$^1$\footnote[3]{takashi.toma@tum.de}
\vspace{5mm}

{\it%
{$^1$ Physik-Department T30d, Technische Universit\"at M\"unchen,\\ James-Franck-Stra\ss{}e, D-85748 Garching, Germany \\}
{$^2$ School of Physics, Korea Institute for Advanced Study, Seoul 02455, South Korea\\}
}

\vspace{8mm}

\abstract{We consider a scenario where the dark matter candidate is a sterile neutrino with sizable self-interactions, described by a dimension-six operator, and with negligible interactions with the Standard Model. The relic abundance is set by the freeze-out of 4-to-2 annihilations in the dark sector plasma	and reproduces the observed dark matter abundance when the sterile neutrino mass is in the range $\sim 500\,{\rm keV} - 20\,\,{\rm GeV}$. The feeble interactions with the Standard Model may lead to observable signals from dark matter decay. Furthermore, the self-interactions can affect the formation of small scale structures. 
We also implement this mechanism in a concrete model where the sterile neutrino self-interaction is due to the exchange of a singlet scalar, and we discuss the relevance of the scalar portal interactions for constructing a complete thermal history of the dark sector.}
\end{center}
\end{titlepage}

\renewcommand{\thefootnote}{\arabic{footnote}}
\newcommand{\bhline}[1]{\noalign{\hrule height #1}}
\newcommand{\bvline}[1]{\vrule width #1}

\setcounter{footnote}{0}

\setcounter{page}{1}

\section{Introduction}

One of the simplest extensions of the Standard Model consists in adding to the matter content a spin 1/2 Majorana fermion, singlet under the Standard Model gauge group, usually denominated sterile neutrino, right-handed neutrino or simply fermion singlet (for reviews, see \cite{Adhikari:2016bei,Abazajian:2017tcc,Kusenko:2009up}). The symmetries of the model allow a Majorana mass term for the sterile neutrino, with size which is a priori unrelated to the electroweak symmetry breaking scale and which can range between 0 and the Planck scale, as well as a Yukawa interaction of the sterile neutrino with the left-handed lepton doublets and the Higgs doublet. This model, in particular, leads to non-vanishing active neutrino masses after electroweak symmetry breaking through the renown seesaw mechanism 
\cite{Minkowski:1977sc,Mohapatra:1979ia,Yanagida:1979as,GellMann:1980vs,Schechter:1980gr}. 

The sterile neutrino has been advocated as a dark matter candidate. The Yukawa interaction leads to the production of sterile neutrinos in the very early Universe from interactions with the plasma of Standard Model particles, as well as to the decay into Standard Model particles. Notably, there are regions of the parameter space where the sterile neutrino is cosmologically long lived and has an abundance equal to the observed dark matter abundance~\cite{Dodelson:1993je}. Nevertheless, current upper bounds on the strength of the Yukawa coupling from X-ray observations rule out this framework as the only mechanism of sterile neutrino dark matter production~\cite{Horiuchi:2013noa,Malyshev:2014xqa}. On the other hand, the production rate could be enhanced in the presence of a lepton asymmetry~\cite{Shi:1998km}, or by further extensions of the model {\it e.g.} by the decay of a scalar singlet which couples to the sterile neutrinos~\cite{Kusenko:2006rh,Petraki:2007gq,Merle:2013wta,Konig:2016dzg}. Due to the smaller Yukawa couplings necessary to reproduce the correct dark matter abundance in these frameworks, the limits from X-ray observations are accordingly relaxed, although it has been argued that these scenarios are in tension with limits from structure formation from the Lyman-$\alpha$ forest and from Milky-Way satellite counts~\cite{Schneider:2016uqi,Baur:2017stq} (see, however,~\cite{Garzilli:2015iwa,2017MNRAS.468.4285L}).
These constraints, on the other hand, could be somewhat alleviated by thermalizing the sterile neutrinos, \textit{e.g.}\ through a coupling to a self-interacting scalar \cite{Hansen:2017rxr}.

In this paper we present a mechanism of sterile neutrino production which is realized even when all portal interactions with the visible sector are negligibly small, thus opening new regions of the sterile neutrino dark matter parameter space. Our framework requires the sterile neutrino to have sizable self-interactions, such that their initial population thermalized in a dark sector plasma at some very early stage of the history of the Universe. We show that the self-interactions induce the freeze-out of the sterile neutrino abundance through 4-to-2 annihilations, and we argue that, for appropriate parameters, the relic population of sterile neutrinos can account for all the dark matter of the Universe. Furthermore, the self-interactions could potentially alleviate the small scale problems of the cold dark matter paradigm (see {\it e.g.} \cite{Tulin:2017ara}). Opening the portal interactions to the Standard Model, which in this framework are no longer directly constrained by the requirement of generating the observed dark matter abundance, allows for additional tests of the model from the decay of the sterile neutrino dark matter.

This framework is constructed along the lines of Strongly Interacting
Massive Particles (SIMPs) as dark matter candidates. General
considerations of the freeze-out of SIMPs via 3-to-2 and 4-to-2
annihilations were discussed
in~\cite{Carlson:1992fn,Hochberg:2014dra}. Various models of SIMP dark
matter have been proposed, where the dark matter candidate is a
pseudoscalar~\cite{Hochberg:2014kqa,Hochberg:2015vrg}, a scalar~\cite{Bernal:2015bla,Choi:2015bya,Ho:2017fte,Bernal:2015xba,Heikinheimo:2016yds,Bernal:2017mqb,Heikinheimo:2017ofk}, a  spin 1/2 Dirac fermion~\cite{Heikinheimo:2016yds,Heikinheimo:2017ofk,Bernal:2017mqb}, a spin 1 vector boson~\cite{Bernal:2015ova,Heikinheimo:2017ofk,Heikinheimo:2018duk} or a
spin 2 boson~\cite{Chu:2017msm}.  Here, we consider for the first time spin 1/2 Majorana fermions as SIMP dark matter. In this case, Lorentz invariance forbids the 3-to-2 annihilation, such that the freeze-out occurs via 4-to-2 annihilations. Furthermore, the Pauli exclusion principle requires the annihilation cross-section to be $d$-wave suppressed, regardless of the nature of the self-interaction.

The paper is organized as follows. In Section \ref{sec:SIMPs} we introduce the framework where sterile neutrinos are SIMP dark matter candidates, with interaction described by a dimension six operator, and we discuss its phenomenology. In Section \ref{sec:toy} we extend the model to include a scalar field which is responsible for the self interactions, and we discuss in detail the role of the scalar field in the cosmology, concretely in populating the dark sector plasma with sterile neutrinos. Finally, we present our conclusions in Section \ref{sec:conclusions}.


\section{Sterile neutrinos as SIMP dark matter candidates}
\label{sec:SIMPs}

The sterile neutrino, that we denote by $\chi$, is a Majorana fermion, singlet under the Standard Model gauge group. Then, the part of the Lagrangian involving only the sterile neutrino field, including operators up to dimension six, reads:
\begin{equation}
\mathcal{L}_{\rm DM}=\frac{1}{2}\overline{\chi^c}i\slashed{\partial}\chi
-\frac{1}{2} m_\chi\overline{\chi^c}\chi +\frac{1}{4!\Lambda^2}(\overline{\chi^c}\chi)(\overline{\chi^c}\chi) +{\rm h.c.}\;.
\label{eq:Lagrangian}
\end{equation}
Besides, there exist portal interaction terms between the visible and the dark sectors, such as the Yukawa coupling $y_\chi  {\overline L} \widetilde H \chi$, where $L$ is a lepton doublet and $\widetilde H= i \tau_2 H^*$, with $H$ the Standard Model Higgs doublet. We will assume in what follows that all portal interactions with the Standard Model have negligible strength, such that they play no role in dark matter production and ensure its stability on  cosmological time scales.

\begin{figure*}[t]
	\begin{center}
		\includegraphics[scale=0.6]{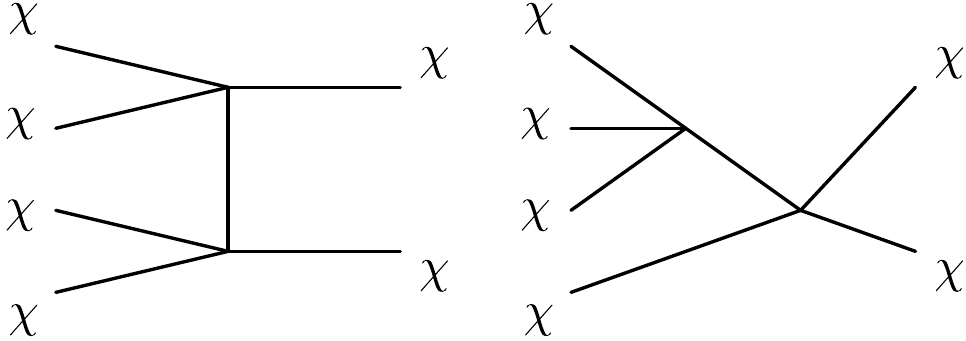}
		\caption{Topologies for the $4$-to-$2$ annihilation process $\chi\chi\chi\chi\to\chi\chi$ in the effective theory of dark matter self-interactions.}
		\label{fig:diagram_eff}
	\end{center} 
\end{figure*}

The dimension-six self-interaction term induces dark matter $2$-to-$2$ scatterings, with a cross section
\begin{align}
\sigma_{2\to2}=\frac{1}{72 \pi }\frac{m_\chi^2}{\Lambda^4}\;,
\end{align}
and which is constrained to be $\sigma_{2\to 2}/m_\chi\lesssim 1\,{\rm cm}^2/{\rm g}$ from Bullet Cluster observations~\cite{Randall:2007ph}; this constraint translates into the lower limit $\Lambda \gtrsim 6 \left(\frac{m_\chi}{\mathrm{MeV}}\right)^{1/4}\MeV$. The $2$-to-$2$ process has been invoked to alleviate the small scale problems of the cold dark matter paradigm (see {\it e.g.} \cite{Tulin:2017ara}), however it does not change the particle number and does not play any role in dark matter production. On the other hand, the 4-to-2 process $\chi\chi\chi\chi\rightarrow \chi\chi$, induced by the diagrams shown in Fig.~\ref{fig:diagram_eff},  does change the particle number and determines the epoch at which the sterile neutrinos can no longer be maintained in thermal equilibrium, thus setting their relic abundance.

We assume in what follows that dark matter particles were produced in the early stages of the Universe and that the strength of the dark matter self-interaction was sufficiently large to keep the dark plasma in thermal equilibrium, characterized by the temperature $T^\prime$. Since we assume negligible portal interactions with the Standard Model plasma, the dark sector temperature, $T^\prime$, is in general different from the visible sector temperature, $T$.

The evolution of the dark matter number density, which we denote by $n_\chi$, is dictated by the following Boltzmann equation:
\begin{align}
\label{eq-bnx}
\frac{dn_\chi}{dt}+3Hn_\chi &= -\langle\sigma{v}^3\rangle \left(n_\chi^4-n_\chi^2{n_\chi^\mathrm{eq}}^2\right)\;,
\end{align}
where $H$ is the Hubble parameter, $\langle\sigma{v}^3\rangle$ is the thermally averaged cross section for the 4-to-2 process and $n_\chi^{\rm eq}$ is the dark matter number density in equilibrium, all evaluated at the corresponding cosmic time $t$. The Hubble parameter is given by:
\begin{align}
H^2=\frac{8\pi}{3}G_N \Big(\rho(T)+\rho'(T')\Big)\;,
\end{align}
with $G_N$ the gravitational constant,  while $\rho(T)$ and  $\rho'(T')$ are respectively the energy densities in the visible and dark sectors:
\begin{align}
\rho(T)=g_{\rm eff}(T)\frac{\pi^2}{30} T^4\;,~~~~
\rho'(T')=g'_{\rm eff}(T')\frac{\pi^2}{30} T'^4\;,
\label{eq:energy-densities}
\end{align}
with  $g_{\rm eff}$ and $g_{\rm eff}^\prime$ the effective number of degrees of freedom in the corresponding sector at that cosmic epoch~\cite{Gondolo:1990dk,Husdal:2016haj}. Besides, the dark matter equilibrium number density reads, assuming Maxwell-Boltzmann statistics:
\begin{align}
n_\chi^{\rm eq}= 
\frac{m_\chi^2 T^\prime}{\pi^2}
 K_2\left(\frac{m_\chi}{T^\prime}\right)\;,
\label{eq:nchi_eq}
\end{align}
with $K_n(x)$ the modified Bessel function of the second kind of 
integer order $n$. Finally, the thermally averaged cross section for the 4-to-2 process is defined as~\cite{Choi:2017mkk}
\begin{equation}
\langle\sigma{v}^3\rangle\equiv
\frac{\displaystyle\int
	d^3v_1d^3v_2d^3v_3d^3v_4\left(\sigma{v}^3\right)\delta^3(\vec{v}_1+\vec{v}_2+\vec{v}_3+\vec{v}_4)e^{-\frac{m_\chi}{2T^\prime}\left(v_1^2+v_2^2+v_3^2+v_4^2\right)}}
{\displaystyle\int d^3v_1d^3v_2d^3v_3d^3v_4\delta^3(\vec{v}_1+\vec{v}_2+\vec{v}_3+\vec{v}_4)e^{-\frac{m_\chi}{2T^\prime}\left(v_1^2+v_2^2+v_3^2+v_4^2\right)}}\;,
\end{equation}
where $\vec v_i$ are the velocities of the initial state particles, while the $4$-to-$2$ annihilation cross section can be obtained from the invariant amplitude ${\cal M}$ from
\begin{equation}
\sigma{v}^3=\frac{\sqrt{3}}{6144\pi m_\chi^4}
\left(\frac{\int\overline{|\mathcal{M}|^2}d\Omega}{4\pi}\right)\;,
\end{equation}	
where $d\Omega$ is the solid angle of any of the final state particles.

We have calculated the invariant amplitude from the Feynman diagrams depicted in Fig.~\ref{fig:diagram_eff}, using FeynCalc~\cite{Mertig:1990an, Shtabovenko:2016sxi}. The final expression is rather lengthy and will not be shown here. Expanding in terms of the dark matter velocities $\vec{v}_i$ ($i=1,2,3,4$), one finds that the leading terms are proportional to fourth-order invariants of the velocities ({\it e.g.} $v_i^4$, $v_i^2v_j^2$, $(\vec{v}_i\cdot\vec{v}_j)^2$), namely the cross-section is $d$-wave suppressed. This behavior can be easily understood from the Pauli exclusion principle. Each initial fermionic state is specified by its spin $s_i=\pm 1/2$ and orbital angular momentum
$\ell_i=0,1,...$. In a partial wave expansion, the state where all the particles have $\ell_i=0$ is incompatible with the Pauli exclusion principle, since this would require two of the fermions to have also the same spin quantum number. Then, the lowest order term in the partial wave expansion must contain two fermions with $\ell=0$ and two fermions with $\ell=1$, thus giving the $d$-wave suppression in the cross section.\footnote{The $d$-wave suppression of the 4-to-2 cross section for Majorana fermions could be circumvented if the fermions have an additional quantum number (for instance, if they are charged under an unbroken global $SU(2)$ symmetry), or if they have higher spin.} This argument can be generalized to other number changing processes involving self-interactions of spin 1/2 (or spin 3/2) Dirac or Majorana fermions.

Using the results from \cite{Choi:2017mkk} for the thermal average of invariants involving velocities, we find for the thermally averaged cross section:
\begin{equation}
\langle\sigma{v}^3\rangle=
\frac{\displaystyle 1201}{245760\sqrt{3}\pi \Lambda^8}\frac{T^{\prime 2}}{m_\chi^2}\;.
\label{eq:sigmav3}
\end{equation}
We note that $\langle\sigma{v}^3\rangle$ and $n^{\rm eq}_\chi$ are
functions of the temperature of the dark sector, whereas the Hubble
parameter depends both on $T$ and $T^\prime$. On the other hand, in the frameworks of relevance for our analysis $T^\prime\ll T$ and therefore the $T^\prime$-dependence of the Hubble parameter can
be neglected. 

The dark matter relic abundance can be calculated using the  instantaneous freeze-out approximation, which states that the dark matter yield, defined as the dark matter number density normalized to the entropy density in the visible sector, remains constant after freeze-out, defined by the time when the rate for the 4-to-2 process equals the Hubble parameter:
\begin{align}
\Gamma_{4\to2}(T_{\rm fo}^\prime)=H(T_{\rm fo})\;,
\label{eq:FO_condition}
\end{align}
where the rate is given by
\begin{align}
\Gamma_{4\to 2}(T^\prime_{\rm fo})=\langle\sigma v^3\rangle(T^\prime_{\rm fo})\, n_\chi^3(T^\prime_{\rm fo})=\langle\sigma v^3\rangle\Big|_{T^\prime=m_\chi}
\left(\frac{T'_{\rm fo}}{m_\chi}\right)^{2} n^3_\chi(T^\prime_{\rm fo})\;.
\label{eq:rate}
\end{align}
Here, we have used the fact that the thermally averaged cross section is $d$-wave suppressed, and hence proportional to  $T'^2$. We remark that by expanding the cross section in powers of the velocity we implicitly require that the dark matter freezes-out while already non-relativistic, which in turn requires a sufficiently low freeze-out temperature. To ensure the validity of our approximations we will conservatively require $T'_{\rm fo}\lesssim m_\chi/3$. 

Finally, the dark matter abundance today can be calculated from 
\begin{align}
\Omega_\chi  = \frac{m_\chi s_0}{\rho_c} Y_{\chi,0}\;,
\label{eq:DM_abundance}
\end{align}
where  $s_0=2890 \,{\rm cm}^{-3}$ and $\rho_c=10.54 h^2\,{\rm GeV}/{\rm m}^3$ are the present time visible sector entropy density and critical density, while  $Y_{\chi,0}$ is the present time dark matter yield, which in the instantaneous freeze-out approximation is equal to the dark matter yield at the freeze-out epoch:
\begin{align}
Y_{\chi,0}\simeq Y_{\chi}(T_{\rm fo})=\frac{n_\chi (T^\prime_{\rm fo})}{s(T_{\rm fo})}\;,
\label{eq:yield_today}
\end{align}
where
\begin{align}
s(T)=\frac{2\pi^2}{45}T^3 h_{\rm eff}(T)\;,
\label{eq:entropy-density}
\end{align}
with $h_{\rm eff}$ the number of effective degrees of freedom contributing to the entropy density of the visible sector~\cite{Gondolo:1990dk,Husdal:2016haj}.

From Eqs.~(\ref{eq:FO_condition},\ref{eq:rate},\ref{eq:DM_abundance}), and from requiring that the sterile neutrino relic abundance reproduces the dark matter abundance measured by Planck, $\Omega_{\rm DM} h^2\simeq 0.12$~\cite{Ade:2015xua}, one obtains in the instantaneous freeze-out approximation the following approximate value of the dark sector freeze-out temperature:\footnote{This is an excellent approximation for
	freeze-out via 4-to-2 annihilations, due to the strong suppression of
	the reaction rate by a factor $\langle \sigma v^3\rangle n^3_\chi\sim
	T'^2 Y^3 s^3 \sim T'^2 T^9$ when the dark matter
	is non-relativistic. For comparison, the rate of 2-to-2 annihilations,
	relevant for the freeze-out of Weakly Interacting Massive Particles (WIMPs), is only suppressed by $\langle \sigma v \rangle n_\chi \sim Y s \sim T^{3}$. We have confirmed the goodness of the approximation by solving exactly the Boltzmann equations for some selected values of the parameters, obtaining a result which differs by at most 10\% from the one obtained with the instantaneous freeze-out approximation.}
\begin{align}
\left(\frac{T'_{\rm fo}}{m_\chi}\right)^{-1}\approx 7.7 + \log \left[\left( \frac{m_\chi}{30\, \mathrm{MeV}}\right)
\left(\frac{T_{\rm fo}/T^\prime_{\rm fo}}{10} \right)^{-3}
\left( \frac{ T'_{\rm fo}}{m_\chi}\right)^{-3/2}
\left( \frac{10}{h_\mathrm{eff}(T_\mathrm{fo})} \right)
\right]\;,
\label{eq-xfrelic}
\end{align}	
which depends on the dark matter mass and on the ratio between the temperatures of the visible and dark sectors at freeze-out. 
For $T_{\rm fo}/T^\prime_{\rm fo}>10$ and $m_\chi<100$ GeV , we find $T^\prime_{\rm fo}\gtrsim m_\chi/18$. Besides, we obtain for the thermally averaged cross section at the epoch  $T^\prime=m_\chi$:
\begin{align}
\langle \sigma v^3\rangle\Big|_{T^\prime=m_\chi} = 6.3 \times 10^{9} \GeV^8
\left( \frac{\mathrm{GeV}}{m_\chi} \right)^{4}
\left( \frac{T'_\mathrm{fo}}{m_\chi/10} \right)^{-9}
\left(\frac{T_\mathrm{fo}/T'_\mathrm{fo}}{10} \right)^{-7}
\sqrt{\frac{g_\mathrm{eff}(T_\mathrm{fo})}{10}}
\left( \frac{h_\mathrm{eff}(T_\mathrm{fo})}{10} \right)^{-3}\;.
\label{eq:sigmav3-cosmo}
\end{align}
Finally, substituting Eq.~(\ref{eq:sigmav3}) into Eq.~(\ref{eq:sigmav3-cosmo}) one finds a value for the suppression scale of the dimension six operator given by:
\begin{align}
\Lambda \approx 25 \MeV \left( \frac{m_\chi}{\mathrm{GeV}}  \right)^{1/2} 
\left( \frac{T_{\rm fo}/T'_{\rm fo}}{10}  \right)^{7/8}  
\left( \frac{T^\prime_{\rm fo}}{m_\chi/10}\right)^{9/8}
\left(\frac{g_\mathrm{eff}(T_{\rm fo})}{10}\right)^{-1/16}
\left(\frac{h_\mathrm{eff}(T_{\rm fo})}{10}\right)^{3/8}\;.
\end{align}	

\begin{figure*}[t]
	\begin{center}
		\includegraphics[width=0.7\textwidth]{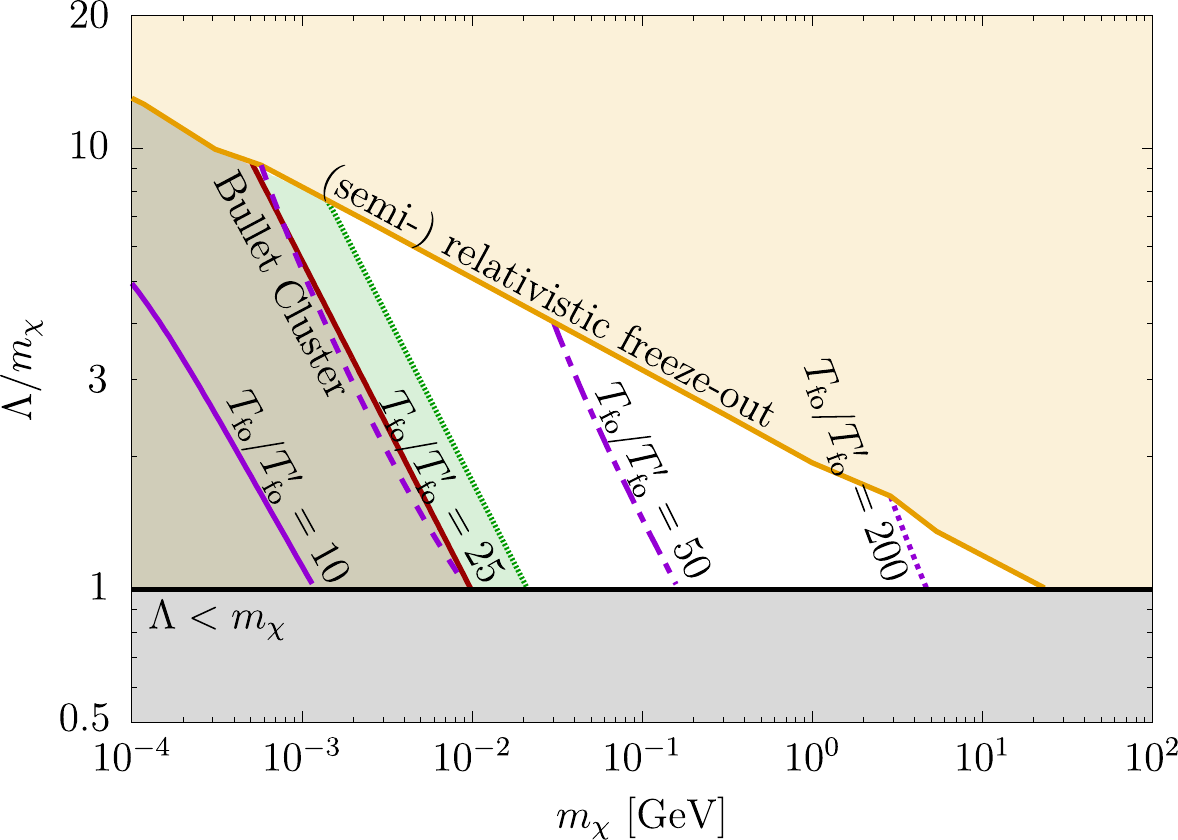}
		\caption{Contours of the temperature ratio between the
	 dark sector and the visible sector at freeze-out leading to the
	 observed dark matter abundance, for given values of the
	 suppression scale of the dimension-six operator and the dark
	 matter mass. The orange region, where the dark matter is
	 semi-relativistic at freeze-out, and the grey region, where the
	 effective theory description is not valid, cannot be analyzed with our set-up. The brown region is excluded by observations of the Bullet Cluster and the green region can
	 potentially alleviate the small scale problems of the cold dark matter paradigm.}
		\label{fig:plot_eff}
	\end{center} 
\end{figure*}

The value of the suppression scale leading to the measured dark matter
abundance, is shown in Fig.~\ref{fig:plot_eff} for different values of the ratio of the freeze-out temperatures $T_{\rm fo}/T'_{\rm fo}=10,25,50,200$.  In our analysis we conservatively disregard the region of the parameter space where  $T'_{\rm fo}>m_\chi/3$, shown as an orange region, to ensure non-relativistic freeze-out, and the region where $\Lambda< m_\chi$, shown as a gray region, to ensure the validity of the effective Lagrangian Eq.~(\ref{eq:Lagrangian}).

The remaining parameter space is constrained by astronomical observations, concretely by the effect of the self-interactions in the Bullet Cluster, which exclude the region of the parameter space where $\sigma_{2\rightarrow 2}/m_\chi> 1\, {\rm cm}^2/{\rm g}$~\cite{Randall:2007ph}, shown as brown in the Figure.
We find that the observed dark matter abundance can be reproduced in
this framework, provided  the dark matter mass is in the range  $
500 \,{\rm keV}\lesssim
m_\chi\lesssim 20$ GeV, which
respectively correspond to a scale of the dimension-six operator,
normalized to the dark matter mass, in the range $10
 \gtrsim \Lambda/m_\chi \gtrsim 1$ and to
a visible-to-hidden sector temperature ratio at freeze-out
$25\lesssim T_{\rm fo}/T'_{\rm fo}\lesssim
400$. 

This model has observable consequences through the dark matter self-interactions. In particular, the self-interaction could provide a solution to the small scale structure problems if $\sigma_{2\rightarrow 2}/m_\chi \gtrsim 0.1 \,{\rm cm}^2/{\rm g}$~\cite{Tulin:2017ara}; this region of the parameter space is shown as green in the Figure and lies in the region where the correct dark matter abundance is set by the 4-to-2 annihilations. 

Additional signals of the model arise when the portal interactions to the Standard Model have non-vanishing strength. More specifically, the Yukawa coupling to the lepton doublet induces dark matter decay, which leads to a flux of cosmic gamma-rays, antimatter particles and neutrinos which, if sufficiently intense, could be disentangled from the background fluxes. Conversely, the non-observation of a statistically significant excess in the measured fluxes can be translated into upper limits on the fundamental parameters of the model. For instance, the non-observation of a significant monoenergetic signal in the gamma-ray sky (see {\it e.g.}~\cite{Essig:2013goa}) can be translated into the following limit on the Yukawa coupling:
\begin{equation}
y_{\chi}\lesssim 10^{-16}\left(\frac{\mathrm{MeV}}{m_\chi}\right)^{3/2}\;, 
\label{eq:Yuk-limit-gamma}
\end{equation}
for $100\,{\rm keV}\lesssim m_\chi \lesssim 100\,{\rm GeV}$. Such small values of the Yukawa coupling could be naturally accommodated in the model by extending its symmetry by an exact, or very mildly broken, discrete $\mathbb{Z}_2$ symmetry, under which the Standard Model fields are even while the sterile neutrino is odd. In the case that the $\mathbb{Z}_2$ symmetry is mildly broken, signals of the model could be observed with future instruments, such as the gamma-ray telescope e-ASTROGAM, which aims to increase the current sensitivity to the decay width for $\chi\to\nu\gamma$ by more than one order of magnitude~\cite{DeAngelis:2016slk}. 
The upper limit Eq.~(\ref{eq:Yuk-limit-gamma}) also implies that the sterile neutrino dark matter candidate contributes negligibly to the active neutrino masses. Concretely, we find $m_\nu\lesssim 10^{-16}\,{\rm eV}  (m_\chi/{\rm MeV})^{-4}$. Neutrino masses could on the other hand be generated by neutrino Yukawa interactions with other fermion singlets, or perhaps by other mechanisms. The observation of dark matter self interactions in astronomical objects, accompanied by the observation of a dark matter decay signal, notably a gamma-ray line in the energy range $250 \,{\rm keV}\lesssim E_\gamma\lesssim 10$ GeV, would provide support to the framework where the dark matter is constituted by sterile neutrinos as SIMPs. 

An important question is whether the strength of the effective self-interaction which is necessary to reproduce the observed dark matter abundance can be naturally accommodated in an ultraviolet complete model. We will address this question in the next section with a simple toy model.

\section{A toy model of dark matter self-interactions}
\label{sec:toy}

We extend the particle content of the dark sector by one scalar singlet $\varphi$, with mass $m_\varphi>m_\chi$, that interacts with the sterile neutrino through a Yukawa coupling of the form:
\begin{equation}
\mathcal{L}_{\rm int }=-\frac{y_\varphi}{2}\varphi\overline{\chi^c}\chi. 
\end{equation}
Then, at the energy scales where the scalar $\varphi$ can be integrated
	out, one recovers the effective interaction in
	Eq.~(\ref{eq:Lagrangian}), with
	$\Lambda=\frac{m_\varphi}{\sqrt{3}y_\varphi}$. In this toy
	model, the Bullet Cluster constraint $\sigma_{2\to
	2}/m_\chi\lesssim 1\,{\rm cm}^2/{\rm g}$ translates into the
	upper limit on the Yukawa coupling $y_\varphi
	\lesssim
	0.1\left(\frac{m_\chi}{\mathrm{MeV}}\right)^{3/4}\left(\frac{m_\varphi}{m_\chi}\right)$.

\begin{figure*}[t]
	\begin{center}
		\includegraphics[scale=0.6]{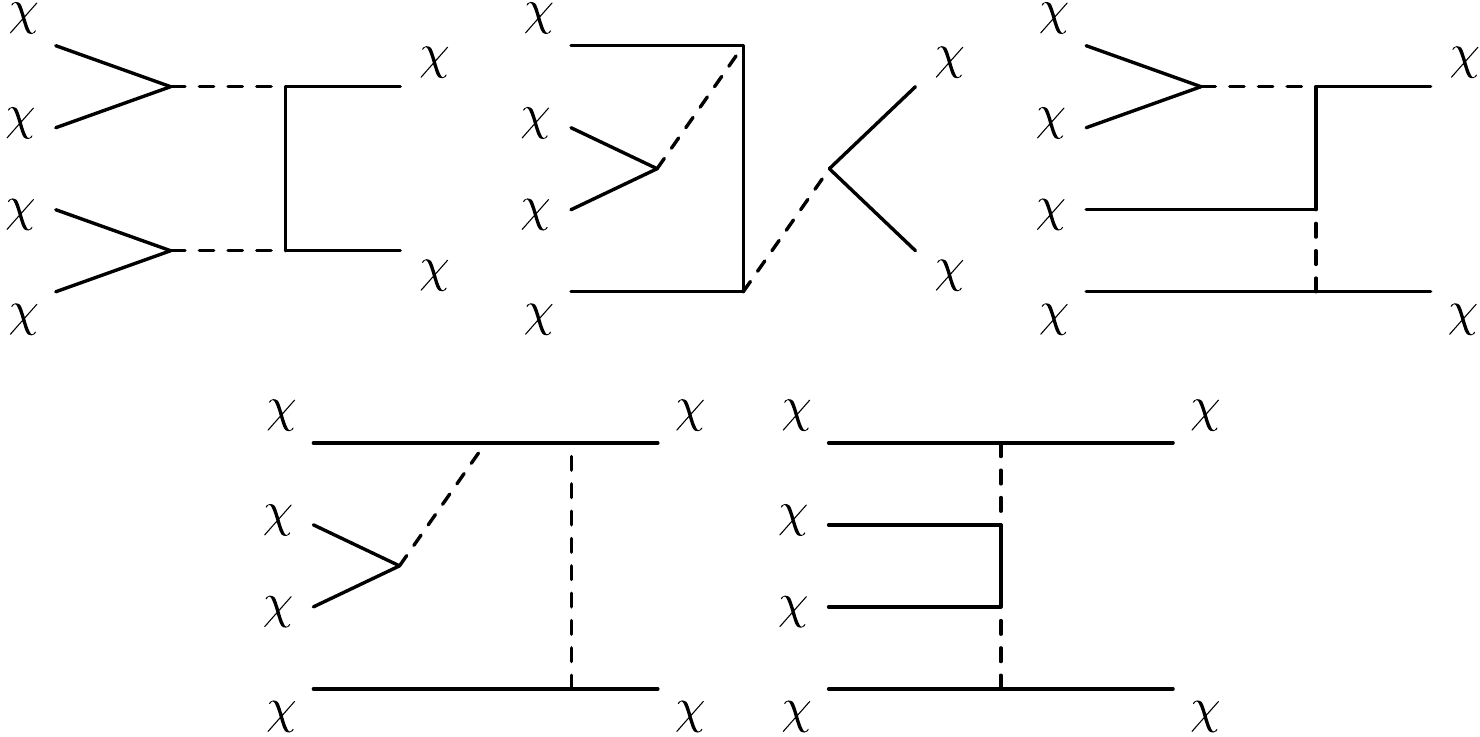}
		\caption{Topologies for the $4$-to-$2$ annihilation process $\chi\chi\chi\chi\to\chi\chi$ in the toy model of dark matter self-interactions through singlet scalar exchange.}
		\label{fig:diagram_toy}
	\end{center} 
\end{figure*}

The process $\chi\chi\chi\chi\to\chi\chi$ is induced in this model by
the diagrams in Fig.~\ref{fig:diagram_toy}. Away from resonances
($2m_\chi\not\approx m_\varphi$ and $4m_\chi\not\approx m_\varphi$), the
thermally averaged annihilation cross section reads:
\begin{equation}
\langle\sigma{v}^3\rangle=
\frac{\displaystyle 27\sqrt{3}y_\varphi^8\sum_{n=0}^{8}a_n\xi^n}{245760\pi
	m_\chi^8(16-\xi)^2(4-\xi)^4(2+\xi)^6{x^\prime}^2},
\label{eq:sigmav3-toy}
\end{equation}
where $\xi\equiv m_\varphi^2/m_\chi^2>4$ and the coefficients $a_n$ are given by
\begin{eqnarray*}
	a_0
	\hspace{-0.15cm}&=&\hspace{-0.15cm}
	2467430400,
	\hspace{0.5cm}
	a_1= - 1648072704,\\
	a_2
	\hspace{-0.15cm}&=&\hspace{-0.15cm}
	491804416,
	\hspace{0.7cm}
	a_3= -25463616,\\
	a_4
	\hspace{-0.15cm}&=&\hspace{-0.15cm}
	4824144,
	\hspace{1.1cm}
	a_5= -1528916,\\
	a_6
	\hspace{-0.15cm}&=&\hspace{-0.15cm}
	473664,
	\hspace{1.3cm}
	a_7= -35259,\\
	a_8
	\hspace{-0.15cm}&=&\hspace{-0.15cm}
	1201.
\end{eqnarray*}

\begin{figure}[t!]
	\begin{center}
		\includegraphics[width=0.49\textwidth]{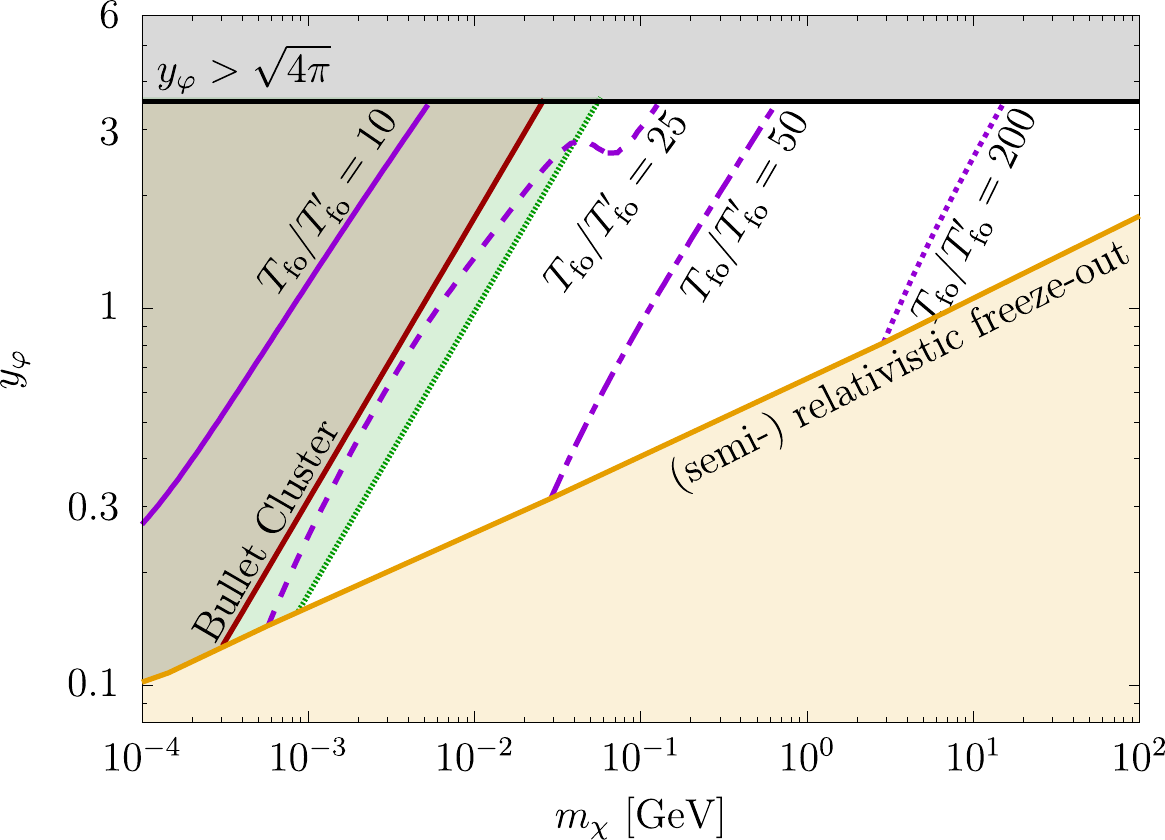}
		\includegraphics[width=0.49\textwidth]{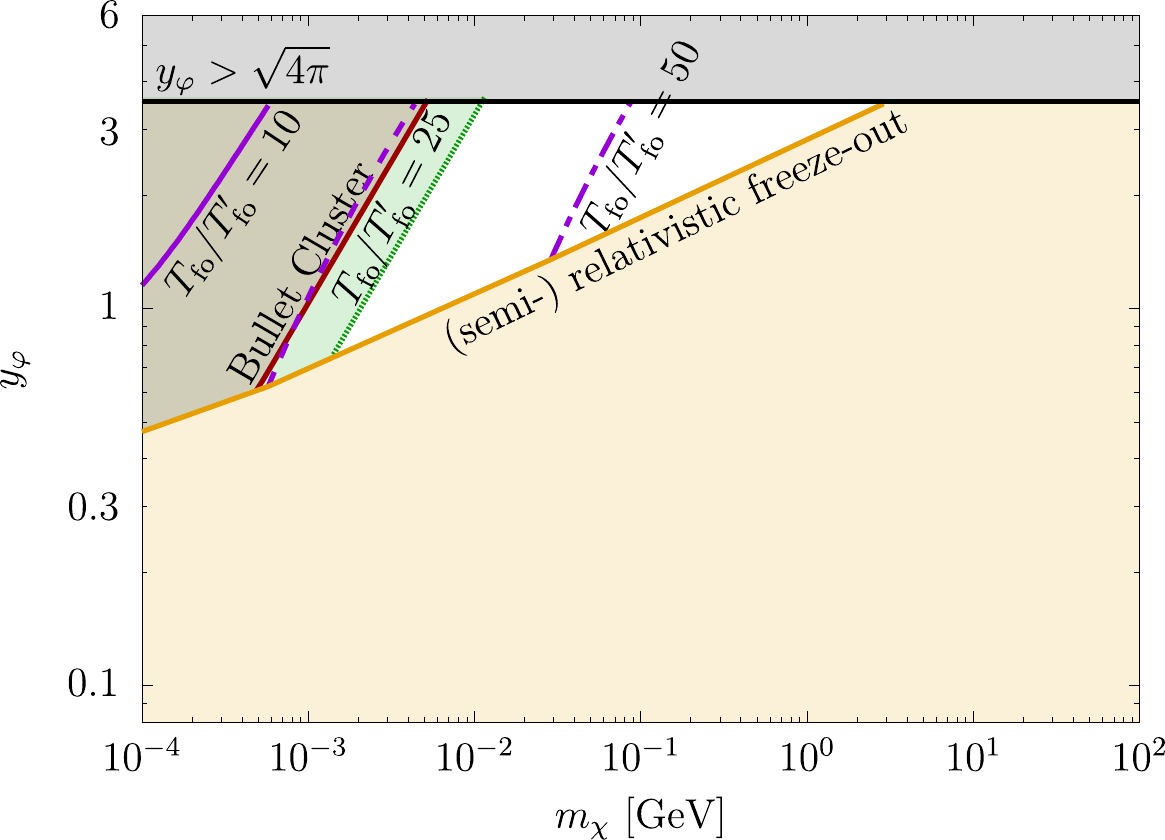}
		\caption{ 
			Contours of the temperature ratio between the dark
			and visible sectors at freeze-out leading to the
			observed dark matter abundance, for given values of the Yukawa
			coupling and the dark matter mass, in a toy model where the
			dark matter self-interaction is due to the exchange of a
			singlet scalar field with mass $m_\varphi=3 m_\chi$ (left
			panel) or $10 m_\chi$ (right panel). In the gray region our perturbative calculation is not valid and cannot be analyzed in our setup. The meaning of the
			other colored regions is as in Fig.~\ref{fig:plot_eff}.}
		\label{fig:toy-model}
	\end{center}
\end{figure}

The value of the Yukawa coupling required to reproduce the observed dark matter abundance can be calculated for a given dark matter mass and mediator mass, and for a given value of the ratio of temperatures at freeze-out of the hidden and visible sectors, using Eqs.~(\ref{eq:sigmav3-toy}) and ~(\ref{eq:sigmav3-cosmo}). The resulting values of the Yukawa coupling as a function of the dark matter mass are shown in Fig.~\ref{fig:toy-model}, for the concrete cases $m_\varphi/m_\chi=3$ (left panel) and 10 (right panel), for $T_{\rm
fo}/T'_{\rm fo}=10, 25, 50, 200$.  The regions of the parameter space
where our approach cannot be applied are indicated in
Fig.~\ref{fig:toy-model} as an orange region and a gray region, and are
bounded by the requirement that the dark matter freezes-out non-relativistically ($T'_{\rm fo}< m_\chi/3$) and that perturbation theory remains valid ($y_\varphi<\sqrt{4\pi}$). Furthermore, the region
indicated in brown implies a strength for the dark matter self-interaction in conflict with observations of the Bullet Cluster.

We find that the observed dark matter abundance can be reproduced by the
thermal freeze-out of sterile neutrinos from the equilibrium density due
to their self-interactions only, if the  sterile neutrino mass is in the range 
$300\,{\rm keV}\lesssim m_\chi\lesssim 3\,{\rm TeV}$ ($500\,{\rm keV}\lesssim m_\chi\lesssim 3\,{\rm GeV}$), the Yukawa coupling is $y_\varphi\gtrsim 0.12$ (0.6)  and the visible-to-hidden sector temperature ratio at freeze-out is $20 \,(24)\lesssim T_{\rm fo}/T'_{\rm fo}\lesssim 2000 \,(200)$ for $m_\varphi/m_\chi=3~(10)$.
We also find points in the parameter space where the small scale problems of the cold dark matter paradigm can be alleviated by dark matter self-interactions, which are shown  in Fig.~\ref{fig:toy-model} as a green region.

We note that for large $m_\varphi/m_\chi$ the effective theory limit is recovered, and therefore the right panel of Fig.~\ref{fig:toy-model} is a just a recast of Fig.~\ref{fig:plot_eff} with $\Lambda=\frac{m_\varphi}{\sqrt{3}y_\varphi}$. In contrast, for moderate $m_\varphi/m_\chi$ the effective Lagrangian does not provide a good description of the model and accordingly the parameters required to reproduce the observed dark matter abundance differ.  Concretely, the dark matter mass window widens both at low and high dark matter masses, and allows a wider range of values for $T_{\rm fo}/T'_{\rm fo}$, as illustrated in Fig.~\ref{fig:plot_eff}, left panel. We also note that for large $m_\varphi/m_\chi$ the 4-to-2 cross section scales as $\sim (y_\varphi m_\chi/m_\varphi)^8$. Therefore, in order to have a sufficiently large annihilation cross section, an increase in  $m_\varphi/m_\chi$ must be compensated by an increase in $y_\varphi$. As a result, for larger and larger  $m_\varphi/m_\chi$ the lower limit on $y_\varphi$ correspondingly increases, eventually entering into conflict with our perturbativity requirement $y_\varphi<\sqrt{4\pi}$, thus setting the upper limit on the singlet scalar mass $m_\varphi\lesssim 55 \, m_\chi$.

The presence of the singlet scalar in the spectrum opens new portals between the visible and the dark sector, through the couplings with the Standard Model Higgs $\mu_{\varphi H}\, \varphi|H|^2$ and $\frac{1}{2} \lambda_{\varphi H}\,\varphi^2|H|^2$. These couplings,  while not directly involved in the freeze-out of the 4-to-2 annihilation process, can be important to address the genesis of the dark sector plasma and hence provide a complete thermal history of the dark matter of our Universe, and in particular of the relation between the temperatures of the visible and dark sector plasmas.\footnote{This is in contrast to WIMP dark matter, where the same interaction that sets the final relic abundance is responsible for thermalizing the dark matter candidate with the Standard Model bath, erasing all previous history of the dark matter population.}
	
More specifically, after electroweak symmetry breaking the particle content of the model contains a real scalar, $h$, which is mainly composed of the Standard Model Higgs boson and has a small component of the scalar singlet $\varphi$, which we parametrize by the mixing angle $\sin\theta$. The Higgs boson is in general in thermal equilibrium with the visible sector plasma and populates the dark sector via freeze-in from the decays $h\rightarrow \varphi\varphi,\chi\chi$, induced by the interaction terms $\frac{1}{\sqrt{2}}\lambda_{\varphi H}\langle H^0 \rangle h \varphi^2$ and $\frac{1}{2} y_\varphi\sin\theta h \overline{\chi^c} \chi$.  The temperature evolution of the dark sector energy density $\rho'$ is described by the following Boltzmann equation~\cite{Chu:2011be}:
\begin{equation}
\frac{d\left(\rho'/\rho\right)}{dT} = - \frac{1}{H T \rho} \Gamma_h m_h n_h(T)\;,
\label{eq:Boltzmann-rho}
\end{equation}
which assumes that all dark sector particles are relativistic while freeze-in production is efficient, namely the epoch $m_h \gtrsim T \gtrsim 10\;\mathrm{GeV}$. In this Boltzmann equation, $\rho$ is the visible sector energy density, given in Eq.~(\ref{eq:energy-densities}),  $n_h$ is the Higgs number density, which at temperatures $T\gtrsim 10\,{\rm GeV}$ is well approximated by its equilibrium value,
\begin{align}
n_h(T)\approx n_h^{\rm eq}(T)= 
\frac{m_h^2 T}{2\pi^2}
K_2\left(\frac{m_h}{T}\right)\;,
\end{align}
and $\Gamma_h$ is the total decay rate into hidden sector particles:
\begin{equation}
\label{eq:gammaHiggs}
\Gamma_h\equiv\Gamma_{h \rightarrow \varphi\varphi}+\Gamma_{h \rightarrow \chi\chi}
= \frac{\lambda_{\varphi H}^2 \langle H^0 \rangle^2}{16 \pi m_h} \sqrt{1-\frac{4 m_\varphi^2}{m_h^2}}
+ \frac{m_h}{16 \pi} y_{\varphi}^2 \sin^2 \theta \left(1-\frac{4 m_\chi^2}{m_h^2}\right)^{3/2}\;.
\end{equation}	
Finally, one can calculate the ratio of dark-to-visible energy densities from solving Eq.~(\ref{eq:Boltzmann-rho}), taking $\rho'/\rho=0$ at high temperature as boundary condition. Correspondingly, the ratio of dark-to-visible sector temperatures can be calculated from Eq.~(\ref{eq:energy-densities}).
 
The temperature ratio is time dependent. It is then convenient to work instead with the entropy ratio, which stays constant over time (as long as there is no entropy production nor energy transfer between the visible and dark sectors) and in particular stays constant during the freeze-out process  \cite{Bernal:2015xba,Carlson:1992fn}. More specifically, we define the entropy ratio as $\zeta\equiv s/s'$, where $s$ is the entropy density of the visible sector, given in Eq.~(\ref{eq:entropy-density}), and $s'$ is the corresponding quantity for the dark sector. We calculate the  value of the entropy ratio by evaluating $\zeta$ at the temperature $T=10\,{\rm GeV}$,  when the freeze-in production is essentially completed. We obtain: 
\begin{align}
\zeta = \left( \frac{T}{T'} \right)^3 \frac{h_\mathrm{eff}(T)}{h'_\mathrm{eff}(T')}\Big|_{T=10\,{\rm GeV}}
\simeq  4.5 \times 10^{5} \left(\frac{\lambda_{\varphi H}}{10^{-10}}\right)^{-3/2} + 7.4 \times 10^{5} \left(\frac{y_\varphi \sin\theta}{10^{-10}}\right)^{-3/2}\;,
\label{eq:EntropyRatio}
\end{align}
where $T'/T$ was calculated using Eq.~(\ref{eq:Boltzmann-rho}), and we have taken $h_{\rm eff}=86.3$ and $h'_\mathrm{eff}=2.75$ at $T=10$ GeV.

Using the fact that the entropy density remains constant, it is straightforward to calculate from Eq.~(\ref{eq:yield_today}) the dark matter yield at the present time in the instantaneous freeze-out approximation. The result is
\begin{align}
Y_{\chi,0}  &= \frac{n^\mathrm{eq}_\chi(T'_{\rm fo})}{s(T_{\rm fo})} 
= \frac{45}{4 \pi^4} \frac{g_\chi}{h_\mathrm{eff}(T_{\rm fo}) } \left( \frac{T'_{\rm fo}}{T_{\rm fo}} \right)^3 \left(\frac{m_\chi}{T'_{\rm fo}}\right)^2 K_2\left(\frac{m_\chi}{T'_{\rm fo}}\right)= \frac{1}{\zeta} F\left(\frac{m_\chi}{T'_{\rm fo}}\right)
\end{align}
with $F(x)\equiv K_2(x)/[x K_1(x) + 4 K_2(x)]$.\footnote{The definition of $F(x)$ follows from the effective number of degrees of freedom contributing to the hidden sector entropy density  at freeze-out
$h_\mathrm{eff}'(T'_{\rm fo}) = \displaystyle{\frac{45}{4\pi^4} g_\chi 
\left[ \left(\frac{m_\chi}{T'_{\rm fo}}\right)^3 K_1\left(\frac{m_\chi}{T'_{\rm fo}}\right) + 4 \left(\frac{m_\chi}{T'_{\rm fo}}\right)^2 K_2\left(\frac{m_\chi}{T'_{\rm fo}}\right) \right]}$ ~\cite{Gondolo:1990dk,Husdal:2016haj}.} Here, $T'_{\rm fo}$ is determined by Eq.~(\ref{eq:FO_condition})  and depends on the value of the thermally averaged cross-section for the 4-to-2 annihilation process, while $T_{\rm fo}$ is related to $T'_{\rm fo}$ by the entropy ratio.  Finally, the sterile neutrino relic abundance follows from Eq.~(\ref{eq:DM_abundance}), and can be readily confronted with the dark matter abundance measured by Planck. 
		
\begin{figure}[t!]
	\begin{center}
		\includegraphics[width=0.49\textwidth]{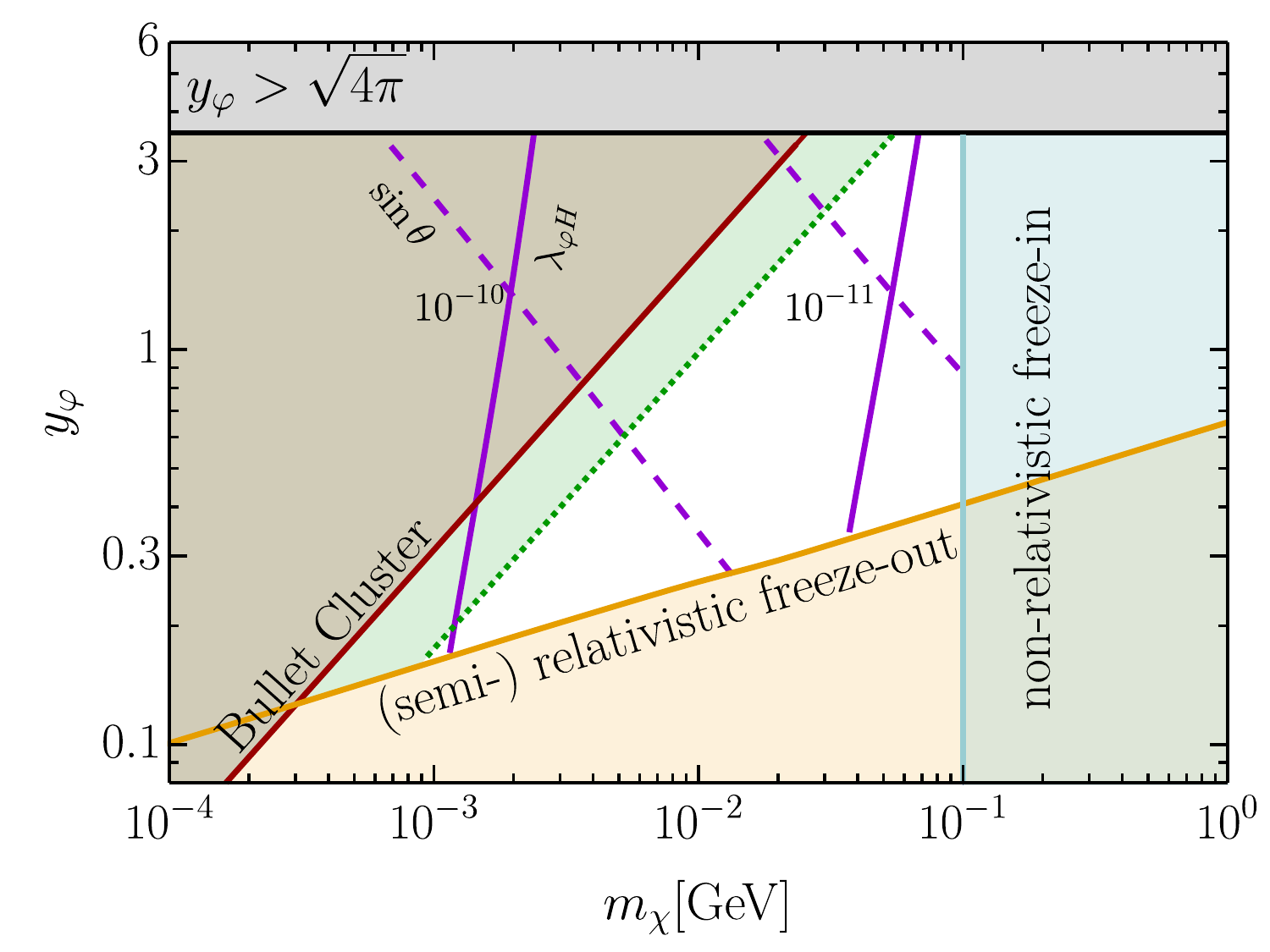}
		\includegraphics[width=0.49\textwidth]{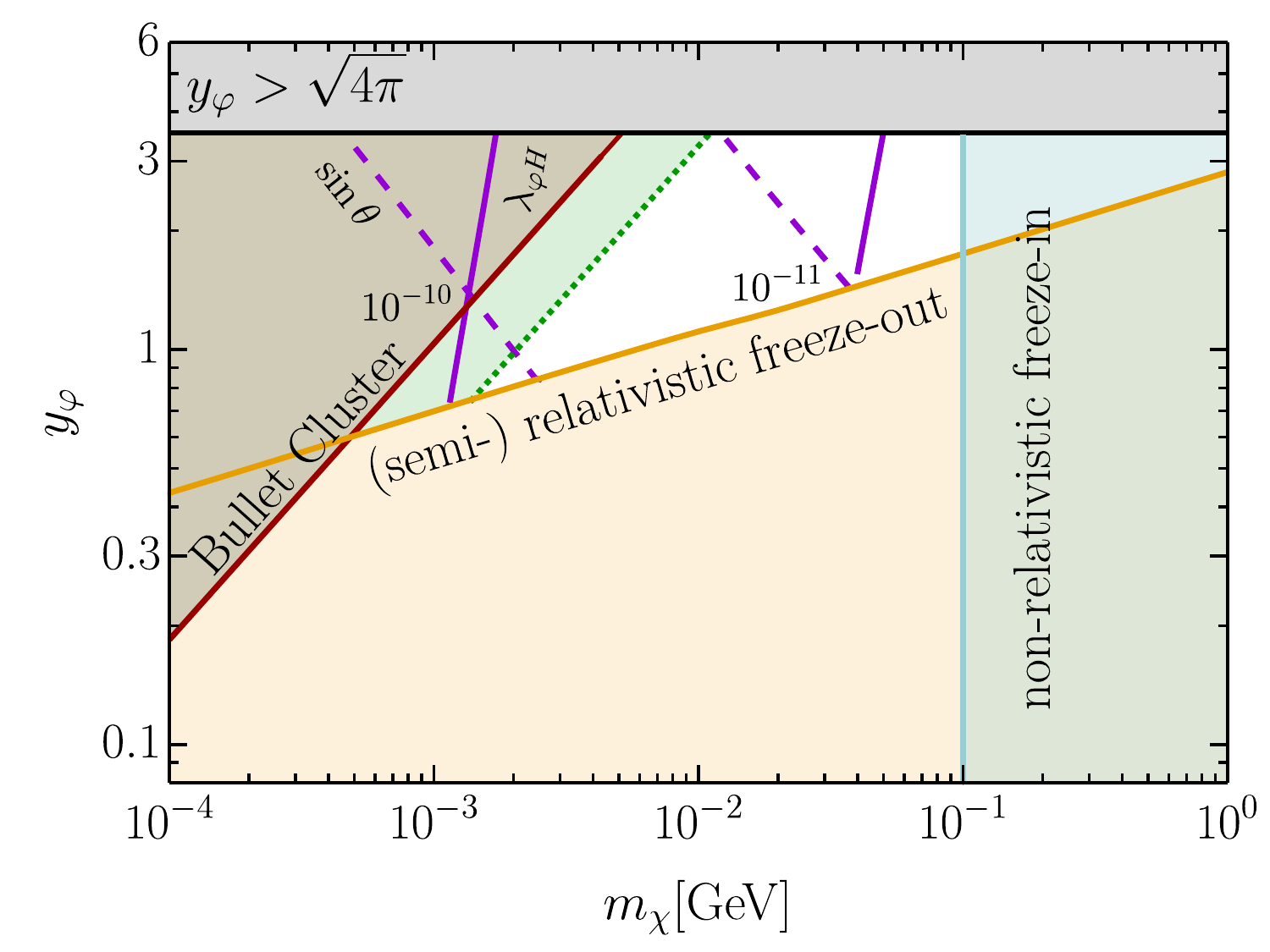}
		\caption{
			Contours of the portal interaction strength leading to the
			observed dark matter abundance for given values of the Yukawa
			coupling and the dark matter mass, in a toy model where the
			dark matter self-interaction is due to the exchange of a
			singlet scalar field with mass $m_\varphi=3 m_\chi$ (left
			panel) or $10 m_\chi$ (right panel). In the light blue region the hidden sector particles are non-relativistic during freeze-in and cannot be analyzed with our approach. The meaning of the other colored regions is as in Fig.~\ref{fig:toy-model}.}
		\label{fig:higgsPortal}
	\end{center}
\end{figure}

We show in Fig.~\ref{fig:higgsPortal} the analog of Fig.~\ref{fig:toy-model}, but plotting contours of the portal coupling $\lambda_{\varphi H}$ or $\sin\theta$ which is necessary to reproduce the observed dark matter abundance via freeze-out of the 4-to-2 annihilations (assuming that only one of them is non-vanishing), in a framework where the dark sector is populated via freeze-in from Higgs decays. In this manner, the viability of one point in the $y_\varphi-m_\chi$ parameter space is dictated by fundamental parameters of the theory (namely $\lambda_{\varphi H}$ or $\sin\theta$), rather than the thermodynamical quantity $T_{\rm fo}/T'_{\rm fo}$. We restrict our analysis to values of the dark matter mass $\leq 0.1$ GeV to ensure that the hidden sector particles are relativistic during freeze-in, so that the Boltzmann equation Eq.~(\ref{eq:Boltzmann-rho}) holds. Under these assumptions, and imposing the restrictions on the strength of the self-interaction from Bullet Cluster observations, we find that reproducing the observed dark matter abundance requires	$3(3)\times 10^{-12}\lesssim \sin\theta \lesssim 3(0.4)\times 10^{-9}$, or $6(6)\times 10^{-12}\lesssim \lambda_{\varphi H}\lesssim 3(2)\times 10^{-10}$, for $m_\varphi/m_\chi=3\,(10)$.

\section{Conclusions}
\label{sec:conclusions}

We have considered a scenario where the dark matter is constituted by Majorana sterile neutrinos with sizable self-interactions, which we describe by a dimension-six operator suppressed by the scale $\Lambda$, and with negligible interactions with the Standard Model particles. We have assumed that at very early times the plasma of sterile neutrinos is in a thermalized state, with a  temperature which is in general different to the temperature of the visible sector plasma, and we have calculated the sterile neutrino relic abundance from the freeze-out of the number changing 4-to-2 annihilation induced by the self-interaction.  

We have shown that, for reasonable values of the model parameters, sterile neutrinos can account for all the dark matter of the Universe. The framework is further constrained by the effect of the self-interactions on astronomical objects, and is partly excluded by observations of the Bullet Cluster. Imposing that the sterile neutrinos freeze-out when already non-relativistic, our results indicate that their mass is  $500 \,{\rm keV}\lesssim m_\chi\lesssim 20$ GeV, that the suppression scale of the dimension-six operator is $5\,{\rm MeV}\lesssim \Lambda \lesssim 20\,{\rm GeV}$, and that the ratio of temperatures between the visible and dark sectors at freeze-out is  $25\lesssim T_{\rm fo}/T'_{\rm fo}\lesssim 400$.
Furthermore, for some choices of the model parameters, the self-interactions can be strong enough to alleviate the small scale problems of the cold dark matter paradigm.

Opening the portal interactions to the Standard Model allows further tests of the model. Specifically, the feeble charged current interaction to the charged leptons would induce dark matter decay. Therefore, the  observation of dark matter self interactions in astronomical objects, accompanied by the observation of a dark matter decay signal, notably a gamma-ray line in the energy range $250 \,{\rm keV}\lesssim E_\gamma\lesssim 10$ GeV, would provide support to our framework.

We have also analyzed the scenario where the self-interaction is induced by the exchange of a scalar singlet field. We have found that reproducing the observed dark matter relic abundance restricts the Yukawa coupling between the singlet scalar and the sterile neutrinos to be larger than 0.12 (0.6) and the sterile neutrino mass to be in the range $300\,{\rm keV}\lesssim m_\chi\lesssim 3\,{\rm TeV}$ ($500\,{\rm keV}\lesssim m_\chi\lesssim 3\,{\rm GeV}$) when the scalar mass is 3 (10) times larger than the sterile neutrino mass. This toy model also contains the necessary elements to construct a complete thermal history of the dark sector, as the portal interactions of the singlet scalar with the Standard Model Higgs can be responsible for populating the dark sector via freeze-in and for determining the temperature of the dark sector plasma. 

\section*{Acknowledgments}

This work has been partially supported by the DFG cluster of excellence EXC 153 ``Origin and Structure of the Universe'' and by the Collaborative Research Center SFB1258. T.T. acknowledges support from JSPS Fellowships for Research Abroad. Numerical computation in this work was carried out at the Yukawa Institute Computer Facility. T.T. thanks Camilo Garcia Cely for fruitful discussions. 

\bibliographystyle{JHEP}
\bibliography{references}

\providecommand{\href}[2]{#2}\begingroup\raggedright\begin{thebibliography}{10}

\bibitem{Adhikari:2016bei}
M.~Drewes et~al., \emph{{A White Paper on keV Sterile Neutrino Dark Matter}},
  \href{https://doi.org/10.1088/1475-7516/2017/01/025}{\emph{JCAP} {\bfseries
  1701} (2017) 025}, [\href{https://arxiv.org/abs/1602.04816}{{\ttfamily
  1602.04816}}].

\bibitem{Abazajian:2017tcc}
K.~N. Abazajian, \emph{{Sterile neutrinos in cosmology}},
  \href{https://doi.org/10.1016/j.physrep.2017.10.003}{\emph{Phys. Rept.}
  {\bfseries 711-712} (2017) 1--28},
  [\href{https://arxiv.org/abs/1705.01837}{{\ttfamily 1705.01837}}].

\bibitem{Kusenko:2009up}
A.~Kusenko, \emph{{Sterile neutrinos: The Dark side of the light fermions}},
  \href{https://doi.org/10.1016/j.physrep.2009.07.004}{\emph{Phys. Rept.}
  {\bfseries 481} (2009) 1--28},
  [\href{https://arxiv.org/abs/0906.2968}{{\ttfamily 0906.2968}}].

\bibitem{Minkowski:1977sc}
P.~Minkowski, \emph{{$\mu \to e\gamma$ at a Rate of One Out of $10^{9}$ Muon
  Decays?}}, \href{https://doi.org/10.1016/0370-2693(77)90435-X}{\emph{Phys.
  Lett.} {\bfseries 67B} (1977) 421--428}.

\bibitem{Mohapatra:1979ia}
R.~N. Mohapatra and G.~Senjanovic, \emph{{Neutrino Mass and Spontaneous Parity
  Violation}}, \href{https://doi.org/10.1103/PhysRevLett.44.912}{\emph{Phys.
  Rev. Lett.} {\bfseries 44} (1980) 912}.

\bibitem{Yanagida:1979as}
T.~Yanagida, \emph{{HORIZONTAL SYMMETRY AND MASSES OF NEUTRINOS}}, {\emph{Conf.
  Proc.} {\bfseries C7902131} (1979) 95--99}.

\bibitem{GellMann:1980vs}
M.~Gell-Mann, P.~Ramond and R.~Slansky, \emph{{Complex Spinors and Unified
  Theories}}, {\emph{Conf. Proc.} {\bfseries C790927} (1979) 315--321},
  [\href{https://arxiv.org/abs/1306.4669}{{\ttfamily 1306.4669}}].

\bibitem{Schechter:1980gr}
J.~Schechter and J.~W.~F. Valle, \emph{{Neutrino Masses in SU(2) x U(1)
  Theories}}, \href{https://doi.org/10.1103/PhysRevD.22.2227}{\emph{Phys. Rev.}
  {\bfseries D22} (1980) 2227}.

\bibitem{Dodelson:1993je}
S.~Dodelson and L.~M. Widrow, \emph{{Sterile-neutrinos as dark matter}},
  \href{https://doi.org/10.1103/PhysRevLett.72.17}{\emph{Phys. Rev. Lett.}
  {\bfseries 72} (1994) 17--20},
  [\href{https://arxiv.org/abs/hep-ph/9303287}{{\ttfamily hep-ph/9303287}}].

\bibitem{Horiuchi:2013noa}
S.~Horiuchi, P.~J. Humphrey, J.~Onorbe, K.~N. Abazajian, M.~Kaplinghat and
  S.~Garrison-Kimmel, \emph{{Sterile neutrino dark matter bounds from galaxies
  of the Local Group}},
  \href{https://doi.org/10.1103/PhysRevD.89.025017}{\emph{Phys. Rev.}
  {\bfseries D89} (2014) 025017},
  [\href{https://arxiv.org/abs/1311.0282}{{\ttfamily 1311.0282}}].

\bibitem{Malyshev:2014xqa}
D.~Malyshev, A.~Neronov and D.~Eckert, \emph{{Constraints on 3.55 keV line
  emission from stacked observations of dwarf spheroidal galaxies}},
  \href{https://doi.org/10.1103/PhysRevD.90.103506}{\emph{Phys. Rev.}
  {\bfseries D90} (2014) 103506},
  [\href{https://arxiv.org/abs/1408.3531}{{\ttfamily 1408.3531}}].

\bibitem{Shi:1998km}
X.-D. Shi and G.~M. Fuller, \emph{{A New dark matter candidate: Nonthermal
  sterile neutrinos}},
  \href{https://doi.org/10.1103/PhysRevLett.82.2832}{\emph{Phys. Rev. Lett.}
  {\bfseries 82} (1999) 2832--2835},
  [\href{https://arxiv.org/abs/astro-ph/9810076}{{\ttfamily
  astro-ph/9810076}}].

\bibitem{Kusenko:2006rh}
A.~Kusenko, \emph{{Sterile neutrinos, dark matter, and the pulsar velocities in
  models with a Higgs singlet}},
  \href{https://doi.org/10.1103/PhysRevLett.97.241301}{\emph{Phys. Rev. Lett.}
  {\bfseries 97} (2006) 241301},
  [\href{https://arxiv.org/abs/hep-ph/0609081}{{\ttfamily hep-ph/0609081}}].

\bibitem{Petraki:2007gq}
K.~Petraki and A.~Kusenko, \emph{{Dark-matter sterile neutrinos in models with
  a gauge singlet in the Higgs sector}},
  \href{https://doi.org/10.1103/PhysRevD.77.065014}{\emph{Phys. Rev.}
  {\bfseries D77} (2008) 065014},
  [\href{https://arxiv.org/abs/0711.4646}{{\ttfamily 0711.4646}}].

\bibitem{Merle:2013wta}
A.~Merle, V.~Niro and D.~Schmidt, \emph{New production mechanism for kev
  sterile neutrino dark matter by decays of frozen-in scalars},
  \href{https://doi.org/10.1088/1475-7516/2014/03/028}{\emph{JCAP} {\bfseries
  1403} (2014) 028}, [\href{https://arxiv.org/abs/1306.3996}{{\ttfamily
  1306.3996}}].

\bibitem{Konig:2016dzg}
J.~K{\"o}nig, A.~Merle and M.~Totzauer, \emph{{keV Sterile Neutrino Dark Matter
  from Singlet Scalar Decays: The Most General Case}},
  \href{https://doi.org/10.1088/1475-7516/2016/11/038}{\emph{JCAP} {\bfseries
  1611} (2016) 038}, [\href{https://arxiv.org/abs/1609.01289}{{\ttfamily
  1609.01289}}].

\bibitem{Schneider:2016uqi}
A.~Schneider, \emph{{Astrophysical constraints on resonantly produced sterile
  neutrino dark matter}},
  \href{https://doi.org/10.1088/1475-7516/2016/04/059}{\emph{JCAP} {\bfseries
  1604} (2016) 059}, [\href{https://arxiv.org/abs/1601.07553}{{\ttfamily
  1601.07553}}].

\bibitem{Baur:2017stq}
J.~Baur, N.~Palanque-Delabrouille, C.~Yeche, A.~Boyarsky, O.~Ruchayskiy,
  \'{E}.~Armengaud et~al., \emph{{Constraints from Ly-$\alpha$ forests on
  non-thermal dark matter including resonantly-produced sterile neutrinos}},
  \href{https://doi.org/10.1088/1475-7516/2017/12/013}{\emph{JCAP} {\bfseries
  1712} (2017) 013}, [\href{https://arxiv.org/abs/1706.03118}{{\ttfamily
  1706.03118}}].

\bibitem{Garzilli:2015iwa}
A.~Garzilli, A.~Boyarsky and O.~Ruchayskiy, \emph{{Cutoff in the Lyman $\alpha$
  forest power spectrum: warm IGM or warm dark matter?}},
  \href{https://doi.org/10.1016/j.physletb.2017.08.022}{\emph{Phys. Lett.}
  {\bfseries B773} (2017) 258--264},
  [\href{https://arxiv.org/abs/1510.07006}{{\ttfamily 1510.07006}}].

\bibitem{2017MNRAS.468.4285L}
M.~R. {Lovell}, S.~{Bose}, A.~{Boyarsky}, R.~A. {Crain}, C.~S. {Frenk}, W.~A.
  {Hellwing} et~al., \emph{{Properties of Local Group galaxies in
  hydrodynamical simulations of sterile neutrino dark matter cosmologies}},
  \href{https://doi.org/10.1093/mnras/stx654}{\emph{MNRAS} {\bfseries 468}
  (July, 2017) 4285--4298}, [\href{https://arxiv.org/abs/1611.00010}{{\ttfamily
  1611.00010}}].

\bibitem{Hansen:2017rxr}
R.~S.~L. Hansen and S.~Vogl, \emph{{Thermalizing sterile neutrino dark
  matter}}, \href{https://doi.org/10.1103/PhysRevLett.119.251305}{\emph{Phys.
  Rev. Lett.} {\bfseries 119} (2017) 251305},
  [\href{https://arxiv.org/abs/1706.02707}{{\ttfamily 1706.02707}}].

\bibitem{Tulin:2017ara}
S.~Tulin and H.-B. Yu, \emph{{Dark Matter Self-interactions and Small Scale
  Structure}},  \href{https://arxiv.org/abs/1705.02358}{{\ttfamily
  1705.02358}}.

\bibitem{Carlson:1992fn}
E.~D. Carlson, M.~E. Machacek and L.~J. Hall, \emph{{Self-interacting dark
  matter}}, \href{https://doi.org/10.1086/171833}{\emph{Astrophys. J.}
  {\bfseries 398} (1992) 43--52}.

\bibitem{Hochberg:2014dra}
Y.~Hochberg, E.~Kuflik, T.~Volansky and J.~G. Wacker, \emph{{Mechanism for
  Thermal Relic Dark Matter of Strongly Interacting Massive Particles}},
  \href{https://doi.org/10.1103/PhysRevLett.113.171301}{\emph{Phys. Rev. Lett.}
  {\bfseries 113} (2014) 171301},
  [\href{https://arxiv.org/abs/1402.5143}{{\ttfamily 1402.5143}}].

\bibitem{Hochberg:2014kqa}
Y.~Hochberg, E.~Kuflik, H.~Murayama, T.~Volansky and J.~G. Wacker, \emph{{Model
  for Thermal Relic Dark Matter of Strongly Interacting Massive Particles}},
  \href{https://doi.org/10.1103/PhysRevLett.115.021301}{\emph{Phys. Rev. Lett.}
  {\bfseries 115} (2015) 021301},
  [\href{https://arxiv.org/abs/1411.3727}{{\ttfamily 1411.3727}}].

\bibitem{Hochberg:2015vrg}
Y.~Hochberg, E.~Kuflik and H.~Murayama, \emph{{SIMP Spectroscopy}},
  \href{https://doi.org/10.1007/JHEP05(2016)090}{\emph{JHEP} {\bfseries 05}
  (2016) 090}, [\href{https://arxiv.org/abs/1512.07917}{{\ttfamily
  1512.07917}}].

\bibitem{Bernal:2015bla}
N.~Bernal, C.~Garcia-Cely and R.~Rosenfeld, \emph{{WIMP and SIMP Dark Matter
  from the Spontaneous Breaking of a Global Group}},
  \href{https://doi.org/10.1088/1475-7516/2015/04/012}{\emph{JCAP} {\bfseries
  1504} (2015) 012}, [\href{https://arxiv.org/abs/1501.01973}{{\ttfamily
  1501.01973}}].

\bibitem{Choi:2015bya}
S.-M. Choi and H.~M. Lee, \emph{{SIMP dark matter with gauged Z$_{3}$
  symmetry}}, \href{https://doi.org/10.1007/JHEP09(2015)063}{\emph{JHEP}
  {\bfseries 09} (2015) 063},
  [\href{https://arxiv.org/abs/1505.00960}{{\ttfamily 1505.00960}}].

\bibitem{Ho:2017fte}
S.-Y. Ho, T.~Toma and K.~Tsumura, \emph{{A Radiative Neutrino Mass Model with
  SIMP Dark Matter}},
  \href{https://doi.org/10.1007/JHEP07(2017)101}{\emph{JHEP} {\bfseries 07}
  (2017) 101}, [\href{https://arxiv.org/abs/1705.00592}{{\ttfamily
  1705.00592}}].

\bibitem{Bernal:2015xba}
N.~Bernal and X.~Chu, \emph{{$\mathbb {Z}_2$ SIMP Dark Matter}},
  \href{https://doi.org/10.1088/1475-7516/2016/01/006}{\emph{JCAP} {\bfseries
  1601} (2016) 006}, [\href{https://arxiv.org/abs/1510.08527}{{\ttfamily
  1510.08527}}].

\bibitem{Heikinheimo:2016yds}
M.~Heikinheimo, T.~Tenkanen, K.~Tuominen and V.~Vaskonen, \emph{{Observational
  Constraints on Decoupled Hidden Sectors}},
  \href{https://doi.org/10.1103/PhysRevD.96.109902,
  10.1103/PhysRevD.94.063506}{\emph{Phys. Rev.} {\bfseries D94} (2016) 063506},
  [\href{https://arxiv.org/abs/1604.02401}{{\ttfamily 1604.02401}}].

\bibitem{Bernal:2017mqb}
N.~Bernal, X.~Chu and J.~Pradler, \emph{{Simply split strongly interacting
  massive particles}},
  \href{https://doi.org/10.1103/PhysRevD.95.115023}{\emph{Phys. Rev.}
  {\bfseries D95} (2017) 115023},
  [\href{https://arxiv.org/abs/1702.04906}{{\ttfamily 1702.04906}}].

\bibitem{Heikinheimo:2017ofk}
M.~Heikinheimo, T.~Tenkanen and K.~Tuominen, \emph{{WIMP miracle of the second
  kind}}, \href{https://doi.org/10.1103/PhysRevD.96.023001}{\emph{Phys. Rev.}
  {\bfseries D96} (2017) 023001},
  [\href{https://arxiv.org/abs/1704.05359}{{\ttfamily 1704.05359}}].

\bibitem{Bernal:2015ova}
N.~Bernal, X.~Chu, C.~Garcia-Cely, T.~Hambye and B.~Zaldivar, \emph{{Production
  Regimes for Self-Interacting Dark Matter}},
  \href{https://doi.org/10.1088/1475-7516/2016/03/018}{\emph{JCAP} {\bfseries
  1603} (2016) 018}, [\href{https://arxiv.org/abs/1510.08063}{{\ttfamily
  1510.08063}}].

\bibitem{Heikinheimo:2018duk}
M.~Heikinheimo, T.~Tenkanen and K.~Tuominen, \emph{{Prospects for indirect
  detection of frozen-in dark matter}},
  \href{https://arxiv.org/abs/1801.03089}{{\ttfamily 1801.03089}}.

\bibitem{Chu:2017msm}
X.~Chu and C.~Garcia-Cely, \emph{{Self-interacting Spin-2 Dark Matter}},
  \href{https://doi.org/10.1103/PhysRevD.96.103519}{\emph{Phys. Rev.}
  {\bfseries D96} (2017) 103519},
  [\href{https://arxiv.org/abs/1708.06764}{{\ttfamily 1708.06764}}].

\bibitem{Randall:2007ph}
S.~W. Randall, M.~Markevitch, D.~Clowe, A.~H. Gonzalez and M.~Bradac,
  \emph{Constraints on the self-interaction cross-section of dark matter from
  numerical simulations of the merging galaxy cluster 1e 0657-56},
  \href{https://doi.org/10.1086/587859}{\emph{Astrophys. J.} {\bfseries 679}
  (2008) 1173--1180}, [\href{https://arxiv.org/abs/0704.0261}{{\ttfamily
  0704.0261}}].

\bibitem{Gondolo:1990dk}
P.~Gondolo and G.~Gelmini, \emph{{Cosmic abundances of stable particles:
  Improved analysis}},
  \href{https://doi.org/10.1016/0550-3213(91)90438-4}{\emph{Nucl. Phys.}
  {\bfseries B360} (1991) 145--179}.

\bibitem{Husdal:2016haj}
L.~Husdal, \emph{{On Effective Degrees of Freedom in the Early Universe}},
  \href{https://doi.org/10.3390/galaxies4040078}{\emph{Galaxies} {\bfseries 4}
  (2016) 78}, [\href{https://arxiv.org/abs/1609.04979}{{\ttfamily
  1609.04979}}].

\bibitem{Choi:2017mkk}
S.-M. Choi, H.~M. Lee and M.-S. Seo, \emph{{Cosmic abundances of SIMP dark
  matter}}, \href{https://doi.org/10.1007/JHEP04(2017)154}{\emph{JHEP}
  {\bfseries 04} (2017) 154},
  [\href{https://arxiv.org/abs/1702.07860}{{\ttfamily 1702.07860}}].

\bibitem{Mertig:1990an}
R.~Mertig, M.~Bohm and A.~Denner, \emph{{FEYN CALC: Computer algebraic
  calculation of Feynman amplitudes}},
  \href{https://doi.org/10.1016/0010-4655(91)90130-D}{\emph{Comput. Phys.
  Commun.} {\bfseries 64} (1991) 345--359}.

\bibitem{Shtabovenko:2016sxi}
V.~Shtabovenko, R.~Mertig and F.~Orellana, \emph{{New Developments in FeynCalc
  9.0}}, \href{https://doi.org/10.1016/j.cpc.2016.06.008}{\emph{Comput. Phys.
  Commun.} {\bfseries 207} (2016) 432--444},
  [\href{https://arxiv.org/abs/1601.01167}{{\ttfamily 1601.01167}}].

\bibitem{Ade:2015xua}
{\scshape Planck} collaboration, P.~A.~R. Ade et~al., \emph{{Planck 2015
  results. XIII. Cosmological parameters}},
  \href{https://doi.org/10.1051/0004-6361/201525830}{\emph{Astron. Astrophys.}
  {\bfseries 594} (2016) A13},
  [\href{https://arxiv.org/abs/1502.01589}{{\ttfamily 1502.01589}}].

\bibitem{Essig:2013goa}
R.~Essig, E.~Kuflik, S.~D. McDermott, T.~Volansky and K.~M. Zurek,
  \emph{{Constraining Light Dark Matter with Diffuse X-Ray and Gamma-Ray
  Observations}}, \href{https://doi.org/10.1007/JHEP11(2013)193}{\emph{JHEP}
  {\bfseries 11} (2013) 193},
  [\href{https://arxiv.org/abs/1309.4091}{{\ttfamily 1309.4091}}].

\bibitem{DeAngelis:2016slk}
{\scshape e-ASTROGAM} collaboration, A.~De~Angelis et~al., \emph{{The
  e-ASTROGAM mission}},
  \href{https://doi.org/10.1007/s10686-017-9533-6}{\emph{Exper. Astron.}
  {\bfseries 44} (2017) 25--82},
  [\href{https://arxiv.org/abs/1611.02232}{{\ttfamily 1611.02232}}].

\bibitem{Chu:2011be}
X.~Chu, T.~Hambye and M.~H.~G. Tytgat, \emph{{The Four Basic Ways of Creating
  Dark Matter Through a Portal}},
  \href{https://doi.org/10.1088/1475-7516/2012/05/034}{\emph{JCAP} {\bfseries
  1205} (2012) 034}, [\href{https://arxiv.org/abs/1112.0493}{{\ttfamily
  1112.0493}}].

\end{thebibliography}\endgroup
\end{document}